\documentclass[runningheads]{llncs}
\usepackage{graphicx}
\usepackage{algorithm}
\usepackage{algorithmic}

\usepackage{multirow}
\usepackage{multicol}
\usepackage{threeparttable}
\usepackage{amsmath}
\usepackage{amssymb}
\usepackage{bbding}
\usepackage{hyperref}

\begin{document}
\title{Multimodal Fusion of Echocardiography and Electronic Health Records for the Detection of Cardiac Amyloidosis}
\titlerunning{Multimodal Fusion of Echo and EHR for the Detection of CA}
%
\author{Zishun Feng\inst{1} \href{mailto:fzs@cs.unc.edu}{\Envelope} \and
Joseph A. Sivak\inst{2} \and
Ashok K. Krishnamurthy\inst{1,3}}
\authorrunning{Z. Feng et al.}
%
\institute{Department of Computer Science, University of North Carolina at Chapel Hill 
\and Division of Cardiology, University of North Carolina at Chapel Hill
\and Renaissance Computing Institute, University of North Carolina at Chapel Hill}
\maketitle              
\begin{abstract}
Cardiac amyloidosis, a rare and highly morbid condition, presents significant challenges for detection through echocardiography. Recently, there has been a surge in proposing machine-learning algorithms to identify cardiac amyloidosis, with the majority being imaging-based deep-learning approaches that require extensive data. In this study, we introduce a novel transformer-based multimodal fusion algorithm that leverages information from both imaging and electronic health records. Specifically, our approach utilizes echocardiography videos from both the parasternal long-axis (PLAX) view and the apical 4-chamber (A4C) view along with patients' demographic data, laboratory tests, and cardiac metrics to predict the probability of cardiac amyloidosis. We evaluated our method using 5-fold cross-validation on a dataset comprising 41 patients and achieved an Area Under the Receiver Operating Characteristic curve (AUROC) of 0.94. The experimental results demonstrate that our approach can achieve competitive results with a significantly smaller dataset compared to prior imaging-based methods that required data from thousands of patients. This underscores the potential of leveraging multimodal data to enhance diagnostic accuracy in the identification of complex cardiac conditions such as cardiac amyloidosis.
\end{abstract}
\section{Introduction}
Cardiac amyloidosis is a highly morbid condition and a significant cause of heart failure. It is most effectively treated when recognized early. While there are several echocardiographic features suggestive of cardiac amyloidosis, these often go unrecognized in many outpatient clinics. Since cardiac amyloidosis is rare, these echocardiographic features might not be checked for every patient. Further, diagnosing cardiac amyloidosis from echocardiography requires significant expertise, but patients with symptoms may not present to expert readers. Typically, suspected cardiac amyloidosis is confirmed by cardiac magnetic resonance (CMR) which is not done routinely as it is an expensive and time-consuming procedure. As a result, there is significant value for methods that can flag patients with suspected cardiac amyloidosis from echocardiography and other routinely collected data such as Electronic Health Records (EHR). 

In clinical practice, the diagnosis of cardiac amyloidosis requires information from multiple aspects, including lab tests, biopsy, and imaging tests. The confirmation of cardiac amyloidosis relies on biopsy, cardiac magnetic resonance, or nuclear imaging, which are costly and risky. On the other hand,  there are lab tests that are fast and low-cost and can provide rich information for the diagnosis. For example, blood and urine tests can detect abnormal proteins that signify amyloidosis. Furthermore, amyloidosis may also affect organ functions like the liver, kidney, and thyroid. Therefore, integrating information from lab tests and other EHR has great potential for the early detection of cardiac amyloidosis.

In this study, we explore different ways of combining echocardiograms and EHR to detect cardiac amyloidosis early. Specifically, we train several models and evaluate their performance against decisions made by experts. The models we consider include: (a) an EHR-only model; (b) a model based on the echocardiogram PLAX view only; (c)  a model based on the echocardiogram A4C view only; (d) a model that combines both the PLAX and A4C views; and finally (e) a multi-modality model that combines the PLAX, A4C view echocardiograms and EHR. To accomplish this, we introduce a transformer-based intermediate fusion model that can achieve an AUROC score of 0.94 on a small dataset with 41 patients (17 in the case group, 24 in the control group).

\section{Related Works}
\subsection{Cardiac Amyloidosis Detection}
Cardiac amyloidosis detection with machine learning is a relatively new approach that has not been studied much. In  \cite{zhang2018fully}, the authors collected PLAX echocardiograms from 852 unique patients (81 in the case group, 771 in the control group) for cardiac amyloidosis detection. The authors trained a VGG network on single-frame echocardiograms and achieved an AUROC score of 0.87. \cite{goto2021artificial} built an AI-based cardiac amyloidosis detection system with both electrocardiograms (ECG) and echocardiograms. The system first used an ECG model to select patients with a high risk of amyloidosis. And a trained echocardiography model was applied to give the final prediction. The echocardiography network was trained on a large dataset with 2828 patients (410 in the case group, 2418 in the control group). The authors trained a 3D CNN model on echocardiography videos and achieved an AUROC score of 0.96. However, the AUROC score of the proposed method in this work will possibly be improved with more training data.

\subsection{Multimodal Learning with Medical Data}
There are numerous modalities in medical data, such as imaging, lab tests, prescriptions, clinical notes, and genomics and \cite{kline2022multimodal,acosta2022multimodal} review multimodal AI methods in healthcare. Some examples include \cite{yap2018} that combined macroscopic and dermoscopic images, and patients' demographic data to classify skin lesions and detect melanoma. The authors used a 2-D convolutional neural network to extract features from images. Then, extracted features were combined with the metadata and sent to fully connected layers for the final prediction. This work is an example of intermediate fusion.  \cite{xu2019multimodal} developed a multimodal ensemble method to fuse medical reports (text data) and EHR (structured data) with imaging data to predict ICD (International Classification of Diseases) codes for diseases.

\section{Method}
In this work, we explore cardiac amyloidosis detection performance of different modalities: 1) with electronic health records only, 2) with imaging data only, and 3) with both EHR and imaging data. For the imaging data-only models, we explore the single-view models with the PLAX view, the A4C view, and a double-view model that combines these two views. For the multimodal classification using both echocardiogram and EHR data, we explore the late fusion and intermediate fusion strategies.

\subsection{Electronic Health Record Classification Model} 
The EHR features contain demographic data and vital signs, cardiac metrics, and lab test values. Given the limited size of the training data (41 patients) and total feature number (166 features), we use early fusion to combine these three features and a logistic regression model to predict the probability of cardiac amyloidosis (Fig. \ref{fig:ch5_inter} a).

\subsection{Echocardiography Video Classification Model} 
\paragraph{Single-View Model}There are two different views of echocardiograms: the PLAX view and the A4C view. We use the spatiotemporal model described in \cite{feng2021two}, which has shown the ability to extract spatiotemporal features from echocardiograms for prediction. The model consists of a convolutional encoder, a bi-directional LSTM with attention, and a feed-forward classifier. The echocardiographic frames are first sent to the convolutional encoder to extract spatial features. The features are then fed into the LSTM with an attention layer to extract temporal features. Finally, the feed-forward classifier predicts the probability of cardiac amyloidosis (Fig. \ref{fig:ch5_inter} b). 

\paragraph{Double-View Model} 
The double-view model combines two single-view models with the late fusion strategy, which has been shown as an effective strategy in video analysis \cite{snoek2005early}. Two single-view models first extract spatiotemporal features from the two views. Then, the two features are concatenated to form the joint feature, which is sent to a feed-forward classifier for the final prediction (Fig. \ref{fig:ch5_inter} c).

\subsection{Mutlimodal Classification Model}
\begin{figure}[h]
\centering
\includegraphics[width=\textwidth]{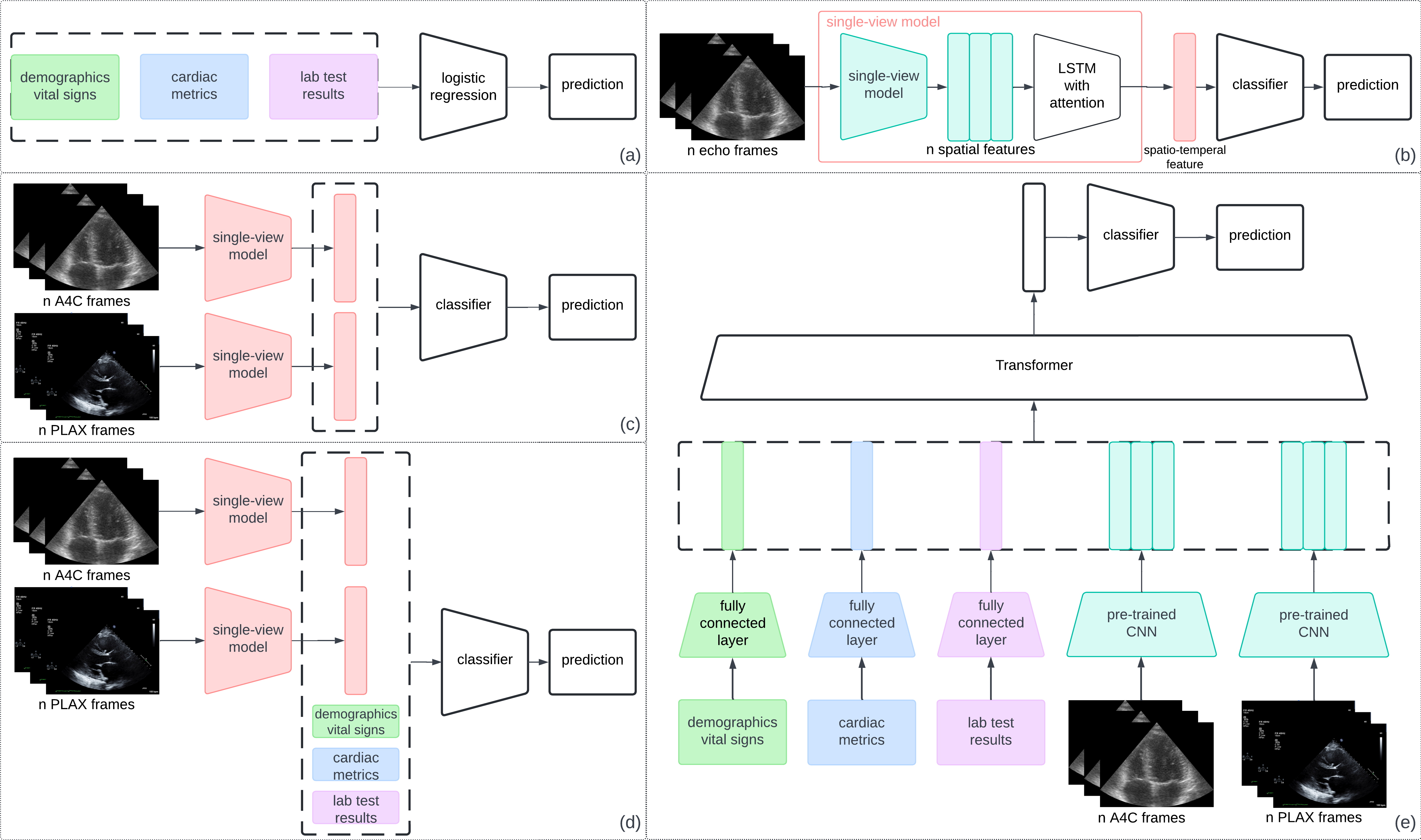}
\caption{Architecture of the intermediate fusion model for multimodal classification. \newline (a) EHR only model. (b) Single-view model. (c) Double-view model with late fusion. (d) Multimodal model with late fusion. (e) Multimodal model with intermediate fusion based on transformer. Components in the dashed box are concatenated. } 
\label{fig:ch5_inter}
\end{figure}

\paragraph{Late Fusion}
The late fusion strategy is one of the most popular fusion strategies in medical applications \cite{kline2022multimodal,huang2020multimodal,feng2022improving}. Late fusion methods first extract one feature from each modality and concatenate these features for the final prediction. Similar to the double-view model, the late fusion multimodal model extracts spatiotemporal features from the PLAX and A4C echocardiograms and aggregates these features with the EHR features. The combined feature is then fed to a fully connected layer-based classifier for prediction (Fig. \ref{fig:ch5_inter} d).

\paragraph{Intermediate Fusion}
We introduce a transformer-based intermediate fusion model. The transformer-based model has shown the capability to learn multimodal features in many previous works \cite{khan2022transformers,li2019visualbert,lu2019vilbert}. The model first maps EHR and imaging data to 512-dimensional features. Specifically, feed-forward layers are used for EHR embeddings; pre-trained convolutional neural networks are used for echocardiogram embeddings. Then, the EHR and imaging data embeddings are fed into a transformer encoder. The transformer layer extracts the joint features, which are sent to fully connected layers to predict the probability of cardiac amyloidosis. Figure \ref{fig:ch5_inter} (e) shows the architecture of the intermediate fusion model.

\subsection{Training Settings}
Since the task is a binary classification task, we chose the binary cross-entropy as the loss function for all the above-mentioned models. All the models were trained using the Adam algorithm with a learning rate of 0.0001 for 20 epochs.

\section{Data}
\subsection{Selected Cohort}
Patients from UNC hospitals and clinics in the Carolina Data Warehouse for Health (CDWH)\footnote{{https://tracs.unc.edu/index.php/services/informatics-and-data-science/cdw-h}} between the ages of 18-99  with a diagnosis of amyloidosis (ICD-10: E85 and ICD-9 277.3) who underwent an echocardiography procedure within a 6-month time period prior to a cardiac magnetic resonance (CMR)  procedure were included. All CMRs were between 01/01/2014 to 01/01/2022. The patients were labeled as positive or negative for cardiac amyloidosis based on the CMR, which is considered the gold standard for detecting the disease. From the EHR we collected demographic data, vital signs, and lab tests from one year before to one year after the diagnosis date (CMR-based diagnosis date). We extracted cardiac metrics from echocardiography imaging reports. After patient selection, the final dataset contains  41 patients with 17 positive and 24 negative samples.
Table \ref{tab:ch5_demo} shows the patients' characteristics.

\begin{table}[htb]
\caption{Demographics and vital signs of patients in the dataset.}
\label{tab:ch5_demo}
\centering
\begin{threeparttable}
\begin{tabular}{l|lll }
\hline
\multirow{2}*{Characteristic}  
 & All & Positive & Negative \\
~ & (n=41)&(n=17)&(n=24)\\
\hline
Age, mean(std), y & 66.1(12.0) & 71.7(11.3) & 62.0(10.8)\\
\hline
Sex: number(percentage) & & & \\
\ \ Female & 14 (34.1\%) &5(29.4\%) &9(37.5\%) \\
\ \ Male & 27 (65.9\%)&12(70.6\%) &15(62.5\%) \\
\hline
Race:number(percentage) & & & \\
\ \ Black or African
American & 20 (48.8\%) &8(47.1\%) &12(50\%) \\
\ \ White& 21 (51.2\%)&9(52.9\%) &12(50\%) \\
\hline
Blood pressure: mean(std), mmHg & & & \\
\ \ Diastolic & 72.5 (15.4) &68.1(10.2) &72.5(17.5) \\
\ \ Systolic & 121.6 (23.9)&112.5(17.6) &128.3(25.6) \\
\hline
Weight: mean(std), kg & 88.9(19.8)&84.1(18.5)&92.3(19.9) \\
\hline
Height: mean(std), m & 1.76(0.1)&1.75(0.1)&1.77(0.1) \\
\hline
BMI: mean(std) & 28.7(5.7)&27.5(4.7)&29.6(6.1) \\ 

\hline
\end{tabular}
\end{threeparttable}
\end{table}

\subsection{Data Preprocessing}
All the EHR data and imaging data contain protected health information (PHI). To protect patients' identifiers, de-identification is always the first step. We first stored all data in a protected environment and removed all the PHI in the data before the development.

The raw electronic health records are usually sparse due to missing values, which cannot be used for prediction directly. Therefore, preprocessing is a necessary step for EHR data. Similarly, the echocardiograms are saved in DICOM (Digital Imaging and Communications in Medicine) files containing static frames and videos from different views. Thus, we need to select the proper echocardiographic videos for development and prediction.

\subsubsection{Preprocessing of Electronic Health Record}
The electronic health records used in this study were extracted from CDWH in the PCORnet \footnote{The National Patient-Centered Clinical Research Network. {https://pcornet.org/}} data model. In this study, we used demographic data, vital signs, and laboratory test results to form the EHR features. The demographic data and vital signs were extracted as the values during the event when the patient underwent the echocardiography procedure. Since the demographic data and vital signs are the basic information that is routinely collected, there are no missing values for these variables and no data imputation was applied. There were 414 different lab tests in the dataset with a large number having very low coverage across patients. We assume that these low-coverage lab tests are not related to the target disease, and only selected the lab tests that happened within 3 months before and 1 month after the echocardiography procedure and had at least a 10 percent coverage across patients. All the missing values were filled with the mean value across all patients for whom the test was available. This resulted in 149 lab tests being included in the dataset for training and testing.

\subsubsection{Preprocessing of Echocardiograms}
The DICOM file for each patient contains echocardiograms from different videos with different settings. To extract the PLAX and A4C views, we used a pre-trained VGG network \cite{simonyan2014very} to identify these two views of echocardiograms in the DICOM file. One of the authors visually checked and confirmed that the correct echocardiogram views were selected by the VGG classification. Further, only the videos with at least 30 frames were selected to ensure that an entire cardiac cycle was included. 

Several cardiac metrics are also included in the DICOM files. The echocardiogram technician manually labeled these cardiac metrics during the procedure and saved the frame with measurements as static images in the DICOM files. To form the cardiac metrics feature vector from the DICOM files we extracted the measurements, including wall thickness, chamber diameter, chamber volumes, stroke volume, ejection fraction, fractional shortening, and sphericity index.

\section{Experiments and Results}
\subsection{Experiment Setting}
In this study, we evaluated multimodal learning from two aspects: 1) modality types and fusion strategy and 2) the contribution of each modality. Limited by the dataset size, we used 5-fold cross-validation for all experiments, and the average score of all 5 folds was reported as the final score. As the task is binary classification, we used accuracy, sensitivity, specificity, and area under the receiver operating characteristic (AUROC) to evaluate the prediction performance.



\subsection{The Impact of Multimodal Learning}
We first evaluated the prediction performance with 1) data from different modalities and 2) different fusion strategies. For the modalities, we tested models with 1) EHR data only, 2) the PLAX view echocardiograms only, 3) the A4C echocardiograms only, 4) the fusion of PLAX and A4C echocardiograms, and 5) the fusion of all three data sources. For the fusion strategies, we tested intermediate fusion and late fusion.

\begin{table}[htb]
\caption{The prediction performance with different modalities and fusion strategies.}
\label{tab:ch5_exp1}
\centering
\begin{threeparttable}
\begin{tabular}{c c|c c c|c c c c}
\hline
&\multirow{2}*{Method}  & \multicolumn{3}{c|}{Data Modality}  & \multirow{2}*{Accuracy}  & \multirow{2}*{Sensitivity}  & \multirow{2}*{Specificity} & \multirow{2}*{AUROC} \\
&~& EHR& PLAX& A4C &~ &~ &~& \\
\hline
1& Logistic regression& \checkmark & & & 0.778 & 0.900 & 0.720 &0.769 \\
\hline
2& CNN+LSTM+Attn& &\checkmark & & 0.851 & 0.897 & 0.818 &0.810 \\
3& CNN+LSTM+Attn& & & \checkmark  & 0.854 & 0.810 & 0.885 &0.772 \\
4& Late fusion& &\checkmark &\checkmark & 0.883 & 0.893 & 0.875 &0.863 \\
\hline
5& Late fusion&\checkmark &\checkmark &\checkmark & 0.912 & 0.894 & 0.925 &0.895 \\
6& Intermediate fusion&\checkmark &\checkmark & \checkmark  & 0.927 & 0.918 & 0.933 &0.941 \\
\hline
\end{tabular}
\end{threeparttable}
\end{table}

The experimental results are shown in Table \ref{tab:ch5_exp1}. From the table, the single modality data (EHR, PLAX, and A4C echocardiograms) can provide good prediction performance. The EHR feature can provide an acceptable predictive power but with a high false positive rate. The imaging data contains more helpful information than the EHR data, and the two echocardiogram views can provide information from different aspects. Thus, combining the two views of echocardiograms can aggregate the information for the two aspects. Line 4 in the table proves that claim; the late fusion of PLAX and A4C views of echocardiograms gives better accuracy and much higher AUROC. Finally, fusing the EHR and imaging data can achieve the highest prediction performance. Moreover, the intermediate fusion can combine the information in an earlier stage and achieve a higher performance than the late fusion method, which aggregates the information from each modality before the final classifier. 

\subsection{Ablation of EHR Components}
From Table \ref{tab:ch5_exp1}, we know that adding the EHR feature to the imaging data can achieve better performance with both intermediate fusion and late fusion. The EHR data consists of three components: 1) demographics and vital signs, 2) cardiac metrics, and 3) lab test results. To investigate the impact of each component in the EHR data, we conducted an ablation study, and  the results are shown in Table \ref{tab:ch5_exp2}

\begin{table}[htb]
\caption{The ablation study of EHR components. Each component is removed from the EHR data to evaluate the impact.}
\label{tab:ch5_exp2}
\centering
\begin{threeparttable}
\begin{tabular}{c|c c c|c c c c}
\hline
\multirow{2}*{Method}  & \multicolumn{3}{c|}{EHR components}  & \multirow{2}*{Accuracy}  & \multirow{2}*{Sensitivity}  & \multirow{2}*{Specificity} & \multirow{2}*{AUROC} \\
~& Demo\&Vital& metrics& Labs &~ &~ &~& \\
\hline
Intermediate &  &\checkmark &\checkmark & 0.912 & 0.882 & 0.933 &0.900 \\
Intermediate & \checkmark& &\checkmark & 0.899 & 0.906 & 0.892 &0.885 \\
Intermediate &\checkmark &\checkmark &  & 0.878 & 0.859 & 0.892 &0.855 \\
\hline
Intermediate &\checkmark &\checkmark & \checkmark  & 0.927 & 0.918 & 0.933 &0.941 \\
\hline
\end{tabular}
\end{threeparttable}
\end{table}

From Table \ref{tab:ch5_exp2}, we can find that dropping the lab test values influenced the performance the most. And the demographic data and vital signs contribute the least to the prediction model. 

\subsection{Contribution to Performance}
To further study the contribution of each modality, we calculated the transformer model's attention weights of each modality. We drew pie charts to show the importance of each modality and each component in the EHR data. Moreover, we also computed the weights of the EHR feature mapping layer to show the contribution of each entry in the EHR data.

\begin{figure}[h]
\centering
\includegraphics[width=\textwidth]{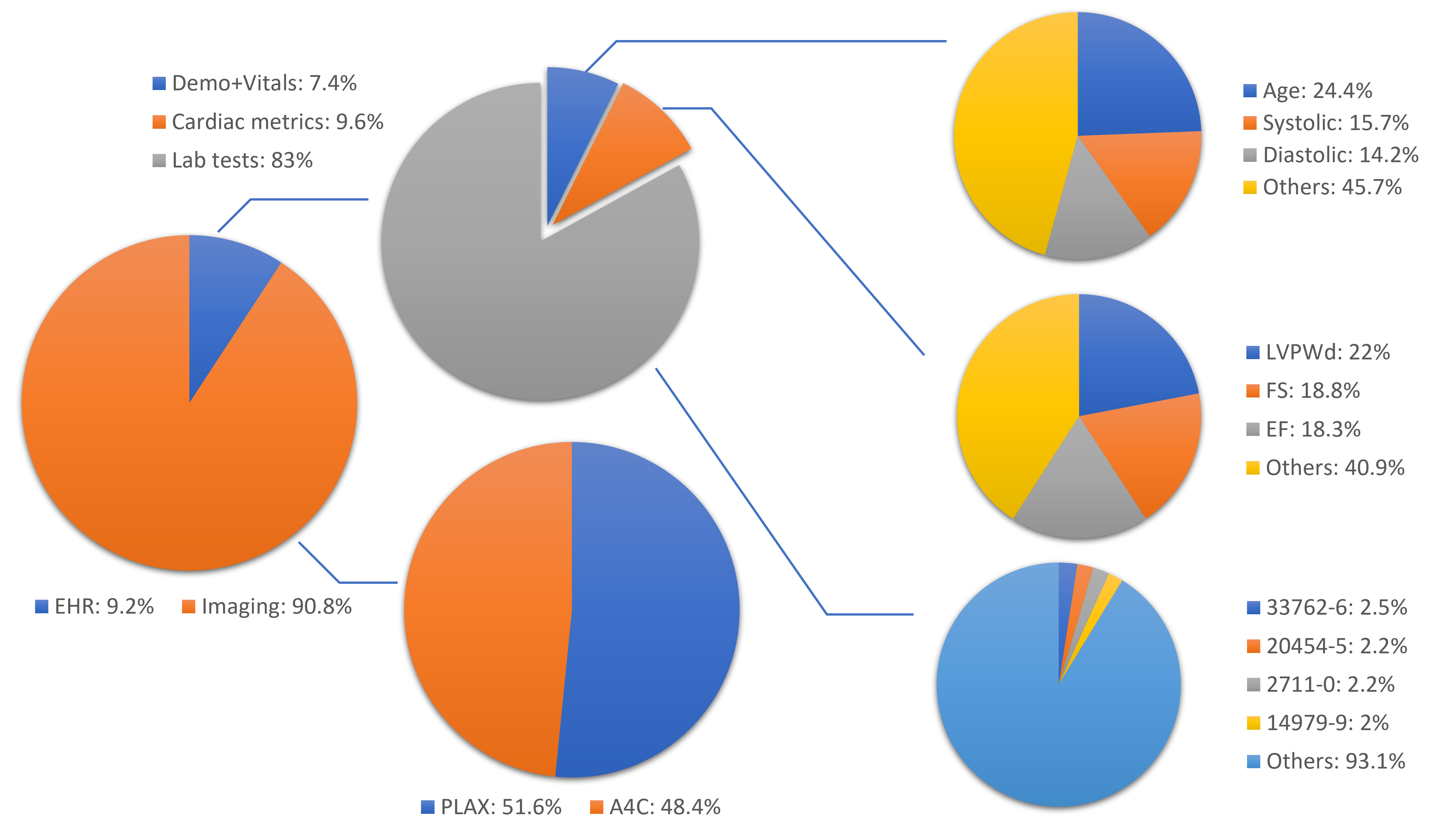}
\caption{Pie charts of feature importance. Left: overall importance. Middle: the importance of EHR and imaging. Right: detailed importance of each entry in the EHR data. The LOINC codes of lab tests are shown in the lower-right corner. } 
\label{fig:ch5_importance}
\end{figure}

From Figure \ref{fig:ch5_importance}, we can find that the imaging data are much more important than the EHR data. The contribution of two echocardiogram views is almost the same. For the EHR data, the lab test result is still the most significant component. Although lab tests are 90\% (149/166) of EHR feature uses, they only contribute 83\% to the EHR performance in prediction, which indicates the proposed model is able to detect non-related lab tests and find other useful non-lab entries. We also find that some lab tests have an importance higher than 2\% (the average is 0.6\%), such as B-type Natriuretic Peptide (BNP) and proteins in urine, which can be considered valuable indicators for cardiac amyloidosis analysis. The top three entries for the cardiac metrics are the wall thickness and indices such as ejection fraction and fractional shortening. For the demographics and vital signs, patients' age and blood pressure contribute the most. 

\subsection{Comparison with Related Studies}
We also compared the proposed multimodal learning-based cardiac amyloidosis detection method with the related studies. Here, we list the method, data type, and prediction performance.

\begin{table}[htb]
\caption{Comparison with other cardiac amyloidosis detection works.}
\label{tab:ch5_comparison}
\centering
\begin{threeparttable}
\begin{tabular}{c|c| c |c| c}
\hline
\multirow{2}*{\ } & \multirow{2}*{Method} & \multirow{2}*{Data Type} & Number of Patients & \multirow{2}*{AUROC} \\
~ & ~ & ~ & (Case/Control) \\
\hline
\multirow{2}*{ \cite{zhang2018fully}} & \multirow{2}*{VGG16} & single PLAX view & 852 & \multirow{2}*{0.87}\\
~ & ~ & echocardiogram frame & (81/771) &~\\
\hline
\multirow{2}*{ \cite{goto2021artificial}} & \multirow{2}*{3D CNN} & A4C view & 2828 & \multirow{2}*{0.96}\\
~ & ~ & echocardiogram video & (410/2418) &~\\
\hline
\multirow{2}*{proposed} & multimodal & EHR & 41 & \multirow{2}*{0.94}\\
~ & learning & PLAX, A4C videos & (17/24) &~\\
\hline
\end{tabular}
\end{threeparttable}
\end{table}

Table \ref{tab:ch5_comparison} shows that the multimodal learning method is able to achieve competitive results even with a small amount of training data. Moreover, by comparing  \cite{zhang2018fully} and the proposed method, utilizing data that contains more information is more helpful than enlarging the size of the dataset. Given more training data, the proposed method will possibly achieve a better performance.

\section{Conclusion}
In this work, we explored the potential of multimodal learning for cardiac amyloidosis detection. We proposed a transformer-based intermediate fusion method for fusing electronic health records and two views of echocardiograms. The model can select valuable identifiers for cardiac amyloidosis diagnosis. Additionally, the model can achieve competitive prediction performance with a small amount of training data.

\section{Acknowledgments}
This work was partially funded by NSF Grant 1633295 BIGDATA: F: Collaborative Research: From Visual Data to Visual Understanding.
%
%
%
\bibliographystyle{splncs04}
\bibliography{references}

\end{document}